\documentclass[english,aps, twocolumn]{revtex4-1}
\usepackage[T1]{fontenc}
\usepackage[latin9]{inputenc}
\usepackage{color}
\usepackage{babel}
\usepackage{amsmath}
\usepackage{amssymb}
\usepackage{graphicx}
\usepackage[unicode=true,pdfusetitle,
 bookmarks=true,bookmarksnumbered=false,bookmarksopen=false,
 breaklinks=false,pdfborder={0 0 0},pdfborderstyle={},backref=false,colorlinks=true]
 {hyperref}
\hypersetup{
 linkcolor=darkblue, citecolor=darkblue, urlcolor=darkblue}

\makeatletter

\newcommand{\lyxmathsym}[1]{\ifmmode\begingroup\def\b@ld{bold}
  \text{\ifx\math@version\b@ld\bfseries\fi#1}\endgroup\else#1\fi}

\usepackage{color}
\usepackage{babel}
\usepackage{breakurl}

\renewcommand{\fnum@figure}{FIG.~\thefigure}
\definecolor{darkblue}{rgb}{0.0, 0.0, 0.55}

\@ifundefined{showcaptionsetup}{}{
 \PassOptionsToPackage{caption=false}{subfig}}

\usepackage[caption=false]{subfig} % use to save caption from breaking when using subfig

\makeatother

\begin{document}
\title{Radiation in equilibrium with plasma and plasma effects on cosmic
microwave background}
\author{Vadim R. Munirov}
\email{vmunirov@pppl.gov}

\affiliation{Princeton Plasma Physics Laboratory, Princeton University, Princeton,
New Jersey 08543, USA}
\affiliation{Department of Astrophysical Sciences, Princeton University, Princeton,
New Jersey 08540, USA}
\author{Nathaniel J. Fisch}
\affiliation{Princeton Plasma Physics Laboratory, Princeton University, Princeton,
New Jersey 08543, USA}
\affiliation{Department of Astrophysical Sciences, Princeton University, Princeton,
New Jersey 08540, USA}
\date{v1 March 13, 2019; v2 June 14, 2019 }
\begin{abstract}
The spectrum of the radiation of a body in equilibrium is given by
Planck's law. In plasma, however, waves below the plasma frequency
cannot propagate; consequently, the equilibrium radiation inside plasma
is necessarily different from the Planck spectrum. We derive, using
three different approaches, the spectrum for the equilibrium radiation
inside plasma. We show that, while plasma effects cannot be realistically
detected with technology available in the near future, there are a
number of quantifiable ways in which plasma affects cosmic microwave
background (CMB) radiation.
\end{abstract}
\maketitle

\section{Introduction}

A system in thermodynamic equilibrium is often said to have a blackbody
radiation spectrum given by Planck's law. However, the Planck spectrum
should be modified within a medium. Indeed, in plasma, for example,
radiation below the plasma frequency $\omega_{p}$ cannot propagate.
Thus, the equilibrium radiation inside plasma is necessarily different
from the Planck spectrum. This paper is dedicated to the investigation
of the equilibrium radiation inside plasma and to the study of the
plasma effects and the possibility of their detection with respect
to one of the most known examples of equilibrium radiation in nature
\textendash{} the cosmic microwave background (CMB).

In Sec.~II we derive the equilibrium spectral energy density of radiation
inside plasma using three different approaches: from the point of
view of photons in plasma treating them as quasiparticles obeying
Bose-Einstein statistics, from the point of view of plasma that generates
electromagnetic fluctuations, and from the point of view of equilibrium
between plasma and external blackbody radiation.

In Sec.~III we consider several questions related to experimental
measurements for stationary and moving observers inside a plasma universe.
We distinguish quantities expressed in terms of frequency from quantities
expressed in terms of wavelength. In a medium with unknown dispersion
we also distinguish energy density from radiation intensity. We also
calculate the Lorentz transformation for a moving observer inside
plasma.

In Sec.~IV we study plasma effects on CMB radiation. We consider
the possibility of experimental detection of static and dynamical
plasma effects on CMB but conclude that these effects cannot be detected
in the next-generation experiments. We show that the static equilibrium
distribution should have been significantly modified during the epoch
of recombination and that this can manifest itself as an extremely
small frequency-dependent chemical potential. We demonstrate that
plasma changes the cosmological redshift and calculate how it distorts
the equilibrium spectrum as the universe expands.

In Sec.~V we consider how plasma modifies the Kompaneets equation
both because of the change in the dispersion relation and because
of coherent scattering in plasma.

In Sec.~VI we consider plasma effects on CMB during and after the
epoch of reionization. We calculate plasma corrections to Compton
$y$-distortion due to the thermal Sunyaev\textendash Zel'dovich effect.
We identify a novel mechanism of magnetic field generation at the
epoch of reionization resulting from conversion of some of the energy
of CMB into the magnetic field. 

We estimate the corrections that plasma effects can bring to other
expected CMB distortions for Planck's telescope \citep{Planck2010}
and SKA-LOW \citep{Dewdney2013} and conclude that plasma effects
are extremely small, on the order of magnitude of $O\left(\omega_{p}^{2}/\omega^{2}\right)$
in most cases, and thus cannot be realistically detected in the near
future.

\section{Radiation in thermodynamic equilibrium with plasma}

Planck's radiation spectrum is characterized only by one parameter
\textendash{} temperature, and it is nonzero for all frequencies.
In equilibrium plasma, however, radiation with frequencies below the
plasma frequency $\omega_{p}$ cannot propagate. Thus, the spectral
energy density of radiation inside plasma is necessarily different
from radiation in free space. Let us derive this spectrum using three
different approaches.

\subsection{Photons as quasiparticles}

Photons are bosons and so they follow Bose-Einstein statistics that
says that the average number of particles with given energy $\varepsilon$
is proportional to $\left[e^{\left(\varepsilon-\mu\right)/T}-1\right]^{-1}$.
Thus, we can write the photon number $n_{\gamma}$ and energy densities
$u_{\gamma}$ as

\begin{equation}
n_{\gamma}=\frac{1}{V}\sum_{\mathbf{k}}\frac{g_{\gamma}}{e^{\frac{\hbar\omega\left(\mathbf{k}\right)-\mu}{T}}-1},
\end{equation}

\begin{equation}
u_{\gamma}=\frac{1}{V}\sum_{\mathbf{k}}\frac{g_{\gamma}\hbar\omega\left(\mathbf{k}\right)}{e^{\frac{\hbar\omega\left(\mathbf{k}\right)-\mu}{T}}-1}.\label{eq:E_general}
\end{equation}

Planck's law follows from these equations if the following assumptions
are employed: $g_{\gamma}=2$, corresponding to two polarizations
of electromagnetic waves; $\mu=0$, corresponding to zero chemical
potential of photons that can be freely absorbed and emitted; non-dispersive
light in vacuum with $\omega=kc$; and substitution of summation with
integration $\left(1/V\right)\sum_{\mathbf{k}}\rightarrow\int d^{3}\mathbf{k}/\left(2\pi\right)^{3}$.
Then Eq.~(\ref{eq:E_general}) yields:

\begin{equation}
u_{Planck}=\frac{\hbar}{\pi^{2}c^{3}}\int_{0}^{\infty}\frac{\omega^{3}}{e^{\frac{\hbar\omega}{T}}-1}d\omega.
\end{equation}

Throughout the paper we will use $u_{\omega}=du/d\omega$ for energy
density per $d\omega$, which for blackbody Planck's radiation we
will denote as $u_{bb}$. We will also use $I_{\omega}$ for intensity
per $d\omega$ defined as $I_{\omega}=v_{gr}u_{\omega}$, where $v_{gr}=\partial\omega/\partial k$
is the group velocity, and $I_{bb}$ for intensity per $d\omega$
for blackbody radiation.

Equation~(\ref{eq:E_general}) shows that the radiation in thermodynamic
equilibrium with plasma or any other matter can be different from
Planck's law for three reasons. First, because of dispersion of waves
in matter, only certain waves with certain frequencies $\omega=\omega\left(\mathbf{k}\right)$
and, as a consequence, energy $\varepsilon\left(\mathbf{k}\right)=\hbar\omega\left(\mathbf{k}\right)$
can propagate in the medium for a given $\mathbf{k}$. Second, there
is nonzero chemical potential $\mu\neq0$. Though it is often approximated
that light has zero chemical potential (for example, Refs.~\citep{Huang1987,Greiner1995}),
it is not the case in general \citep{Wurfel1982,Herrmann2005}. Chemical
potential is related to constraints on the number of particles and
such situations can be realized, for example, in semiconductors \citep{Wurfel1982,Herrmann2005,Ries1991,Markvart2008},
where the number of photons is related to the number of electrons
and holes; in plasmas when scattering dominates over absorption and
thus the number of photons conserved \citep{Zeldovich1969}; in dye
filled microcavities \citep{Klaers2010b,Klaers2014}; and in other
systems \citep{Hafezi2015,Wang2018,Meyer2009}. Third, there are geometrical
and finiteness effects restricting the number of available $\mathbf{k}$
modes. The sum over $\mathbf{k}$ in $\sum_{\mathbf{k}}$ in general
should be performed only over certain $\mathbf{k}$'s. For example,
in cavities depending on the size and geometry only certain waves
with given $\mathbf{k}$'s can exist \citep{Reiser2013,Sokolsky2014,McGregor1978}.

Now let us consider the case of infinite plasma. The typical nonrelativistic
plasma has the following dispersion relation for electromagnetic waves:

\begin{equation}
\omega^{2}=\omega_{p}^{2}+k^{2}c^{2},\label{eq:dispersion}
\end{equation}

\noindent where the plasma frequency is defined through the sum over
the species of charged particles in plasma: $\omega_{p}^{2}=\sum_{s}4\pi n_{s}e_{s}^{2}/m_{s}$.
Since photon energy is $\varepsilon=\hbar\omega$, we can rewrite
Eq.~(\ref{eq:dispersion}) as

\begin{equation}
\varepsilon^{2}=m_{\gamma}^{2}c^{4}+p_{\gamma}^{2}c^{2},
\end{equation}

\noindent i.e., in plasma, a photon behaves as a relativistic massive
particle with mass $m_{\gamma}=\hbar\omega_{p}/c^{2}$ and momentum
$p_{\gamma}=\hbar k$. It is interesting to calculate the density
of electrons for which the effective photon mass equals the electron
rest mass. It happens for electron density $n_{e}\approx(r_{e}\lambda_{C}^{2})^{-3}\approx10^{31}\:\textrm{\textrm{c\ensuremath{\textrm{m}^{\textrm{-3}}}}}$
(corresponding plasma frequency is about $10^{20}\:\textrm{\ensuremath{\textrm{s}^{-1}}}$),
where $r_{e}$ is the classical electron radius and $\lambda_{C}$
is the Compton wavelength. Thus, for extremely high density plasmas,
photons can be expected to behave similarly to massive elementary
particles like electrons.

Going from summation to integration and introducing dimensionless
parameters $a=\hbar\omega_{p}/T$ and $\mu^{\prime}=\mu/T$, we can
write the photon number and energy densities in terms of the normalized
wave vector $y=\hbar kc/T$ as: 

\begin{equation}
n_{\gamma}=\frac{T^{3}}{\pi^{2}\hbar^{3}c^{3}}\int_{0}^{\infty}\frac{y^{2}}{e^{\sqrt{y^{2}+a^{2}}-\mu^{\prime}}-1}dy,
\end{equation}

\begin{equation}
u_{\gamma}=\frac{T^{4}}{\pi^{2}\hbar^{3}c^{3}}\int_{0}^{\infty}\frac{\sqrt{y^{2}+a^{2}}y^{2}}{e^{\sqrt{y^{2}+a^{2}}-\mu^{\prime}}-1}dy,
\end{equation}

\noindent and in terms of the normalized frequency $x=\hbar\omega/T$
as:

\begin{equation}
n_{\gamma}=\frac{T^{3}}{\pi^{2}\hbar^{3}c^{3}}\int_{a}^{\infty}\frac{x^{2}}{e^{x-\mu^{\prime}}-1}dx,
\end{equation}

\begin{equation}
u_{\gamma}=\frac{T^{4}}{\pi^{2}\hbar^{3}c^{3}}\int_{a}^{\infty}\frac{x^{2}\sqrt{x^{2}-a^{2}}}{e^{x-\mu^{\prime}}-1}dx.\label{eq:E_x}
\end{equation}

Thus, the radiation distribution inside plasma is described by three
parameters: temperature $T$, chemical potential $\mu$, and parameter
$a=\hbar\omega_{p}/T$. The parameter $a$ is a measure of the density.

Using the expansion

\begin{equation}
\frac{1}{e^{\sqrt{y^{2}+a^{2}}-\mu^{\prime}}-1}=\sum_{l=1}^{\infty}e^{-l\left(\sqrt{y^{2}+a^{2}}-\mu^{\prime}\right)},
\end{equation}

\noindent substituting $y=a\sinh\theta$, and employing the integral
representation for the modified Bessel functions of the second kind,
we can get the total number and energy densities:

\begin{equation}
n_{\gamma}=\frac{T^{3}}{\pi^{2}\hbar^{3}c^{3}}\sum_{l=1}^{\infty}\frac{e^{l\mu^{\prime}}}{l^{3}}\left(la\right)^{2}K_{2}\left(la\right),\label{eq:n_ph}
\end{equation}

\begin{equation}
u_{\gamma}=\frac{T^{4}}{\pi^{2}\hbar^{3}c^{3}}\sum_{l=1}^{\infty}\frac{e^{l\mu^{\prime}}}{l^{4}}\left[\left(la\right)^{3}K_{1}\left(la\right)+3\left(la\right)^{2}K_{2}\left(la\right)\right].\label{eq:E_ph}
\end{equation}

\noindent Similar expressions were obtained in Refs.~\citep{Bannur2006,Trigger2007,Triger2010,Medvedev1999,Tsintsadze1996,Mati2019}.

\begin{figure}
\includegraphics[width=1\columnwidth]{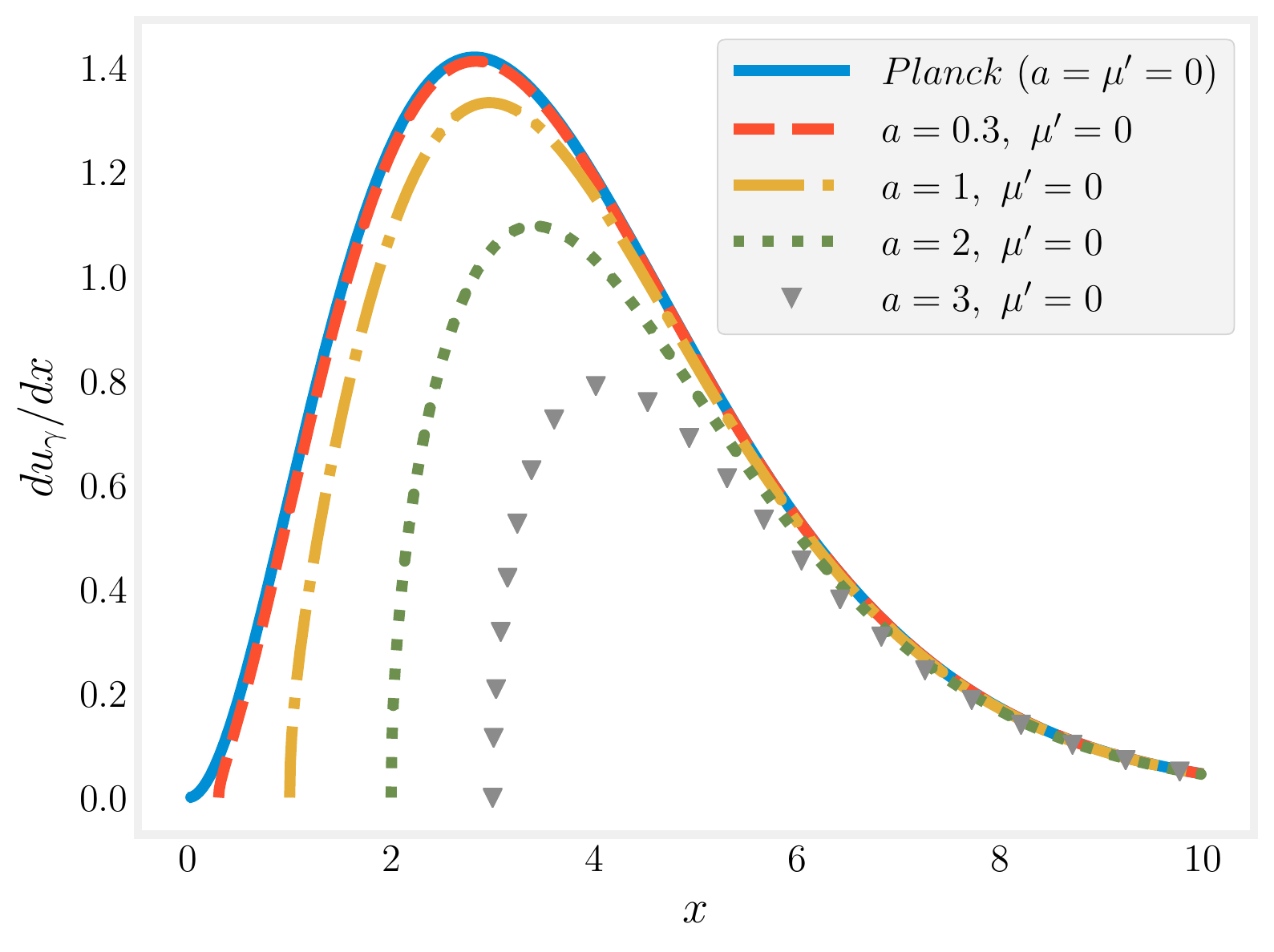} \caption{\label{fig01}Photon energy density in the units of $T^{4}/\pi^{2}\hbar^{3}c^{3}$
versus normalized frequency $x=\hbar\omega/T$ for different values
of parameter $a$ and $\mu^{\prime}=0$ . Blackbody Planck density
corresponding to $a=\mu^{\prime}=0$ is also shown for comparison. }
\end{figure}

Figure~\ref{fig01} shows spectral energy density $du_{\gamma}/dx$
in the units of $T^{4}/\pi^{2}\hbar^{3}c^{3}$ as a function of normalized
frequency $x=\hbar\omega/T$ for several values of parameter $a=\hbar\omega_{p}/T$
and zero chemical potential. We see the truncation of the spectrum
for frequencies below the plasma frequency as well as an overall decrease
of the radiation density with growth of $a$. We also see that for
$a\gtrsim1$ the radiation density starts to significantly deviate
from the Planck distribution, which corresponds to $a=\mu^{\prime}=0$. 

\begin{figure}
\includegraphics[width=1\columnwidth]{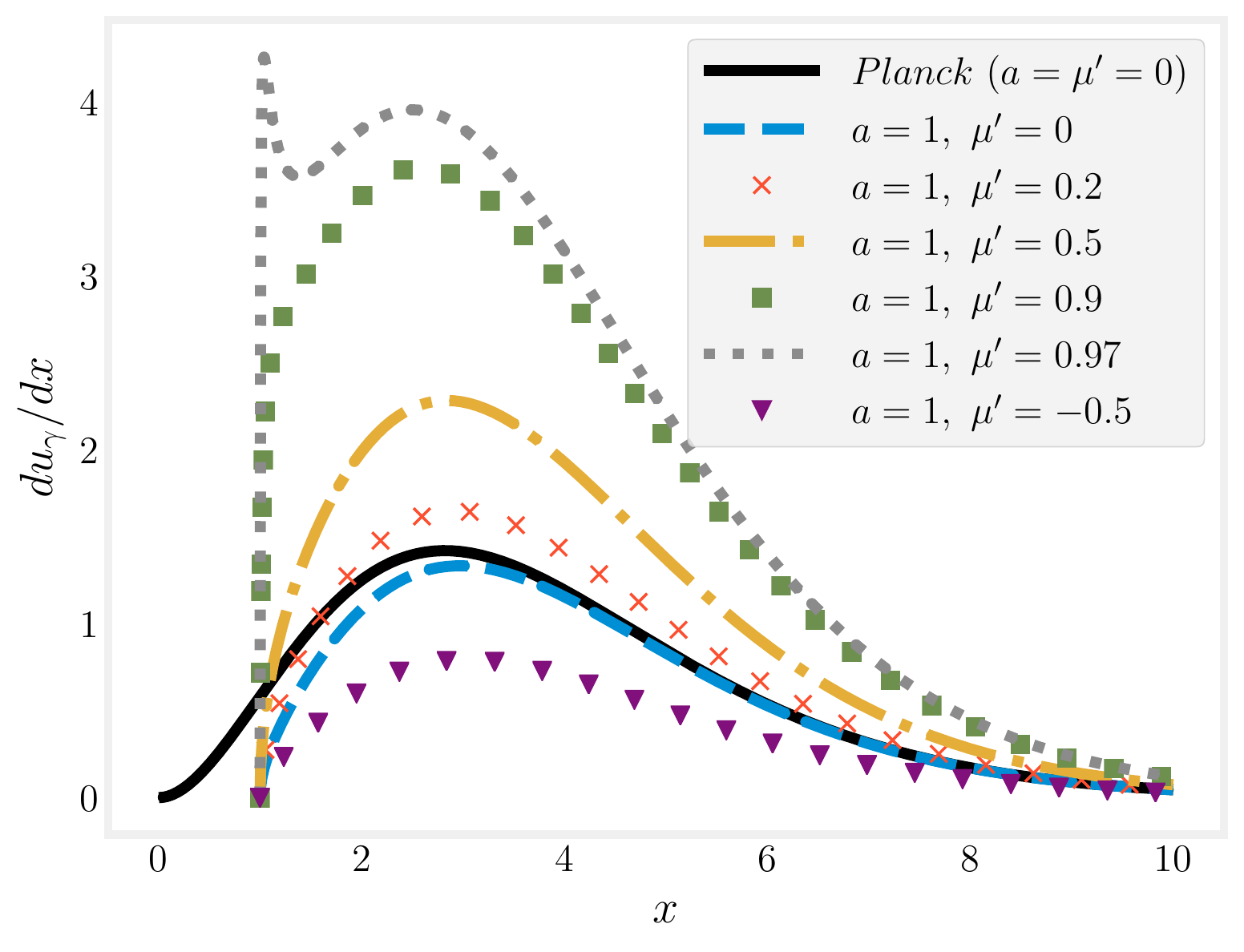} \caption{\label{fig02}Photon energy density in the units of $T^{4}/\pi^{2}\hbar^{3}c^{3}$
versus normalized frequency $x=\hbar\omega/T$ for different values
of chemical potential $\mu^{\prime}$ and $a=1$. Blackbody Planck
density corresponding to $a=\mu^{\prime}=0$ is also shown for comparison. }
\end{figure}

Figure~\ref{fig02} shows spectral energy density $du_{\gamma}/dx$
in the units of $T^{4}/\pi^{2}\hbar^{3}c^{3}$ as a function of normalized
frequency $x=\hbar\omega/T$ for fixed parameter $a=1$, but different
chemical potentials. Despite the occasional claim that bosons can
have only zero or negative chemical potential (for example, Refs.~\citep{Greiner1995,Chen2003}),
in fact they can have positive chemical potential too; the only restriction
is that the smallest energy level is larger than zero and the corresponding
integrals converge. We see that the chemical potential can affect
the distribution significantly. However, we do not address the question
of whether a particular chemical potential can be realistically achieved
in any physical system. Note also that Fig.~\ref{fig02} does not
correspond to one system with given $a$ and different chemical potentials,
but to several independent systems with different number of photons.
For a given system with a fixed number of photons, one cannot vary
the chemical potential independent of $T$ and $a$; see Ref.~\citep{Mati2019}
for details.

Figure~\ref{fig02} also suggests that the radiation energy can exceed
Planck's radiation density. However, it does not contradict the often-made
statement that thermal blackbody radiation establishes the upper limit
on the maximum energy emitted for any body at the same temperature
(for example, Refs.~\citep{Bohren1983,Massoud2005}), because Fig.~\ref{fig02}
shows the energy density inside the plasma and not the intensity that
will be emitted by plasma. Moreover, objects for which the absorption
cross section exceeds the geometrical cross section can actually emit
more than blackbody, but this does not contradict standard physics
and fits the generalized form of Kirchhoff's law \citep{Greffet2018}.

\subsection{Fluctuation-dissipation theorem}

So far we have derived properties of equilibrium radiation from the
point of view of photons obeying Bose-Einstein statistics. Since the
radiation is in equilibrium with the matter, the same results should
be obtained by considering oscillating electrons in plasma and calculating
the density of the electromagnetic field generated by them using the
fluctuation-dissipation theorem \citep{Triger2010,Lifshitz1980}.

Following Ref.~\citep{Lifshitz1980}, the spectral energy density
per $d\omega$ in a transparent medium with dielectric function $\varepsilon\left(\omega\right)$
is given by 

\begin{equation}
u_{\gamma}=\int_{0}^{\infty}\frac{1}{8\pi}\left[2\left(E^{2}\right)_{\omega}\frac{\partial\left(\omega\varepsilon\right)}{\partial\omega}+2\left(H^{2}\right)_{\omega}\right]\frac{d\omega}{2\pi},
\end{equation}

\noindent where the fluctuations $\left(E^{2}\right)_{\omega}$ and
$\left(H^{2}\right)_{\omega}$ of the electric and magnetic fields
are defined through

\begin{equation}
\left\langle \mathbf{E}^{2}\right\rangle =\int_{-\infty}^{\infty}\left(E^{2}\right)_{\omega}\frac{d\omega}{2\pi}=\int_{0}^{\infty}2\left(E^{2}\right)_{\omega}\frac{d\omega}{2\pi},
\end{equation}

\begin{equation}
\left\langle \mathbf{H}^{2}\right\rangle =\int_{-\infty}^{\infty}\left(H^{2}\right)_{\omega}\frac{d\omega}{2\pi}=\int_{0}^{\infty}2\left(H^{2}\right)_{\omega}\frac{d\omega}{2\pi}.
\end{equation}

According to Ref.~\citep{Lifshitz1980}, the electromagnetic field
fluctuations can be expressed as

\begin{multline}
\left(E^{2}\right)_{\omega}=\frac{1}{\varepsilon}\left(H^{2}\right)_{\omega}=\frac{2\omega^{2}\hbar\varepsilon^{\frac{1}{2}}}{c^{3}}\coth\frac{\hbar\omega}{2T}\\
=\frac{4\omega^{2}\hbar\varepsilon^{\frac{1}{2}}}{c^{3}}\left(\frac{1}{2}+\frac{1}{e^{\frac{\hbar\omega}{T}}-1}\right),
\end{multline}

\noindent so we have

\begin{equation}
du_{\gamma}=\frac{\hbar\omega^{2}\varepsilon^{\frac{1}{2}}}{2\pi^{2}c^{3}}\left[\frac{\partial\left(\omega\varepsilon\right)}{\partial\omega}+\varepsilon\right]\left(\frac{1}{2}+\frac{1}{e^{\frac{\hbar\omega}{T}}-1}\right)d\omega.
\end{equation}

Using $\varepsilon\left(\omega\right)=1-\omega_{p}^{2}/\omega^{2}$
for plasma and ignoring zero-field fluctuations ($1/2$ term), we
get the same result for the spectral energy density of radiation in
plasma as previously obtained: Eq.~(\ref{eq:E_x}) with $\mu=0$
.

\subsection{Equilibrium with blackbody walls}

\begin{figure}
\includegraphics[scale=0.7]{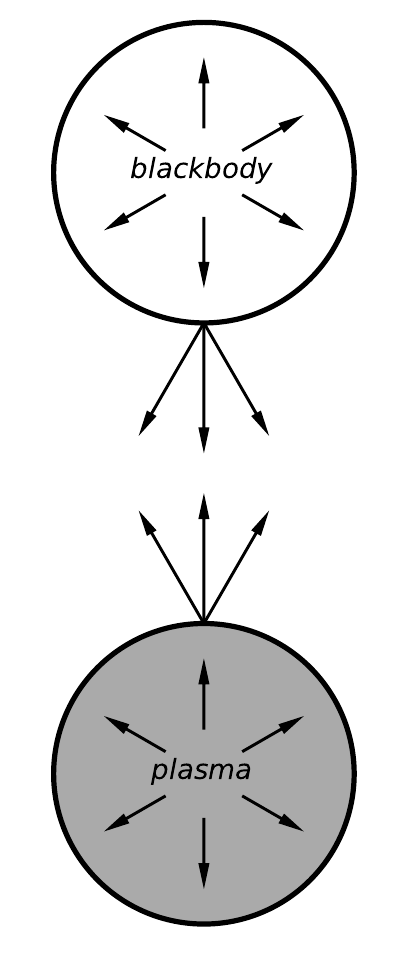} \caption{\label{fig03}Equilibrium between blackbody radiation and plasma,
showing detailed balance among emission, absorption, and reflection. }
\end{figure}

Consider a lossless medium (plasma) surrounded by blackbody walls
at temperature $T$ and being in equilibrium with them (see Fig.~\ref{fig03}).
Let us look at a ray of frequency $\omega_{0}$ and energy $u_{bb}\left(\omega_{0}\right)d\omega_{0}$,
which is emitted by the blackbody walls, and then propagates through
vacuum and impinges on the medium at angle $\theta_{0}$ to its normal.
The wave will experience refraction, such that its frequency remains
the same ($\omega=\omega_{0}$), while its wave vector changes. A
change in the wave vector implies a change in aperture. Specifically,
using Snell's law $n=c/v_{ph}=\sin\theta_{0}/\sin\theta$, the solid
angle $d\Omega=2\pi\sin\theta d\theta$ in plasma can be expressed
through the original solid angle as:

\begin{equation}
d\Omega=\left(\frac{v_{ph}}{c}\right)^{2}\frac{\cos\theta_{0}}{\cos\theta}d\Omega_{0},\label{eq:dsigma}
\end{equation}

\noindent which is the well-known \'etendue conservation condition
\citep{Markvart2008}.

In addition, the group velocity, which determines the energy transport,
changes in the medium. Since in lossless medium the energy must be
conserved, the pulse must be shortened or stretched depending, correspondingly,
on whether the group velocity in the medium is smaller or larger than
in vacuum. 

Finally, only part of the energy of the wave will be transmitted into
the medium, because part of the wave will be reflected. The reflection
coefficient for unpolarized light is $R=\left(R_{s}+R_{p}\right)/2$,
where $R_{s}$ and $R_{p}$ are the reflectances of $s$-polarized
and $p$-polarized electromagnetic waves given by the Fresnel equations.
The transmission coefficient is given by $1-R$.

Taking into account both the change in the aperture and the group
velocity, we can write the energy conservation law:

\begin{multline}
\left(1-R\right)u_{bb}c\cos\theta_{0}d\omega_{0}d\Omega_{0}+Ru_{\omega}v_{gr}\cos\theta d\omega d\Omega\\
=u_{\omega}v_{gr}\cos\theta d\omega d\Omega.\label{eq:trans}
\end{multline}

\noindent Here the first term on the left-hand side is the transmitted
energy, while the second term is the energy of the wave incident on
the medium-vacuum interface at angle $\theta$ from within the medium
and reflected back. Notice that the reflection coefficients coincide
because the reflectance of the light incident from medium 1 on medium
2 at the angle $\theta_{i}$ and refracted into the angle $\theta_{t}$
is the same as the reflectance of the light incident from medium 2
on medium 1 at the angle $\theta_{t}$. Thus, using Eqs.~(\ref{eq:dsigma})
and~(\ref{eq:trans}), we get that the energy density inside the
medium is independent of the reflection coefficient $R$ and is given
by

\begin{equation}
u_{\omega}=\frac{c^{3}}{v_{gr}v_{ph}^{2}}u_{bb}.
\end{equation}

\noindent For plasma $v_{gr}=\partial\omega/\partial k=c\sqrt{1-\omega_{p}^{2}/\omega^{2}}$
, $v_{ph}=\omega/k=c/\sqrt{1-\omega_{p}^{2}/\omega^{2}}$, and using
the Planck energy density, we again get the same result for the electromagnetic
radiation energy density inside plasma as before.

The above result can also be considered through the radiation transfer
equation in the medium \citep{Bekefi1966,Zheleznyakov1996}: 

\begin{eqnarray}
n_{r}^{2}\frac{d}{dl}\left(\frac{I_{\omega}}{n_{r}^{2}}\right) & = & \frac{1}{v_{gr}}\frac{\partial I_{\omega}}{\partial t}+n_{r}^{2}\frac{\partial}{\partial l}\left(\frac{I_{\omega}}{n_{r}^{2}}\right)=\alpha_{\omega}-\mu_{\omega}I_{\omega},
\end{eqnarray}

\noindent where $I_{\omega}$ is the ray intensity per $d\omega$
and is related to the energy density through group velocity as $u_{\omega}=du/d\omega=I_{\omega}/v_{gr}$,
$n_{r}$ is the ray refractive index, which is the usual refractive
index for isotropic medium (see Ref.~\citep{Bekefi1966}), $\alpha_{\omega}$
is emissivity, and $\mu_{\omega}$ is the absorption coefficient (including
scattering).

The condition of transparency of the medium, which corresponds to
$\alpha_{\omega}=\mu_{\omega}=0$ and makes the right-hand side of
the radiation transfer equation zero, results in $I_{\omega}/n^{2}=const$
along the ray. The condition $I_{\omega}/n^{2}=const$ gives $I_{\omega}=n^{2}I_{bb}=\left(1-\omega_{p}^{2}/\omega^{2}\right)I_{bb}$,
i.e., the same radiation energy density inside plasma as before. The
condition $I_{\omega}/n^{2}=const$ also means that $I_{\omega}$
is constant inside plasma as expected for transparent medium. However,
another way to make the right-hand side zero is to have $\alpha_{\omega}/\mu_{\omega}=I_{\omega}$.
In this case $I_{\omega}$ is constant inside the medium as well;
not because the medium does not absorb and emit radiation at all,
but because it does so in a very particular way. This is what actually
happens in plasma, where equilibrium is reached through balance between
emission and absorption (and scattering).

We emphasize that $u_{\omega}$ is the energy density inside the bulk
of the plasma and does not determine the radiation emitted from the
plasma. The radiation leaving the plasma experiences refraction and
lengthening in accordance with the above formulas such that emission
from equilibrium plasma above the plasma frequency is just that of
a blackbody given by the Stefan-Boltzmann law, as expected in equilibrium.
Below the plasma frequency, plasma is a perfect reflector. There must
also be plasma waves along the surface of the boundary. However, the
boundary effects at the plasma-vacuum interface are a separate and
intricate topic beyond the scope of the considerations here.

\section{Measuring plasma spectrum by stationary and moving observers}

Let us consider the experimental detection of radiation \textit{inside}
plasma (or, in general, any kind of dispersive medium). Imagine we
have two observers: one is in a plasma-filled universe in thermodynamic
equilibrium at temperature $T$ and the other is immersed in blackbody
radiation of the same temperature $T$ in vacuum. Imagine both have
identical devices manufactured and calibrated in vacuum that allow
them measure electromagnetic radiation spectrum. What would the observers
actually measure and how should these results be interpreted? Will
they be able to see the difference?

First, we should say that it is intensity and not energy density that
is being measured. In vacuum they are related through the speed of
light, but in the medium they are related through the group velocity,
which depends on unknown properties of the medium. Second, in vacuum
the wavelength and frequency are related through the speed of light
($\omega=2\pi c/\lambda$), so any physical quantity expressed in
terms of frequency can be immediately expressed in terms of wavelength;
for example, if we know intensity per frequency $I_{\omega}$, then
we immediately know intensity per wavelength $I_{\lambda}$. In the
medium frequency and wavelength are related through $\omega=2\pi c/\lambda n\left(\omega\right)$,
i.e., the conversion depends on unknown properties of the medium.
Thus, we should be specific whether the observer in plasma measures
at the same wavelength or at the same frequency as in vacuum.

We can imagine that the observer with a measuring instrument is in
a small vacuum bubble inside the plasma or the measuring instrument
is in direct contact with plasma. In any case, since the light ray
traveling through two different mediums keeps its frequency but changes
its wavelength, all the observable quantities should be expressed
in terms of frequency, i.e., the observer will measure intensity per
frequency $I_{\omega}$. If the bubble is in equilibrium with plasma,
then, from the analysis of the previous section, it is apparent that
the radiation inside the bubble would be that of blackbody and the
device in equilibrium with plasma universe would register just blackbody
radiation spectrum. Thus, the bubble should not be in equilibrium
and, ideally, the measuring instruments should not radiate. This can
be achieved by keeping the temperature of the measuring instruments
close to absolute zero as is done, for example, on Planck's telescope
where the active refrigeration system keeps the HFI detector temperature
at $0.1\:\textrm{K}$ \citep{Planck2010}. Another effect that should
be taken into account when interpreting the experimental results is
that the telescope has inevitable reflections. They can be accounted
for in vacuum, but, in the plasma universe, the reflection coefficient
will be different and generally speaking unknown. Thus, the observer
will measure $\left[1-R\left(\omega\right)\right]I_{\omega}$, where
the reflection coefficient $R\left(\omega\right)$ is not known.

Finally, we address how the spectrum changes for a moving observer.
Since the Doppler shift depends on properties of the medium, the transformation
of the spectrum in the moving reference frame would be different than
that in vacuum. As shown in Ref.~\citep{Bicak1975} for the emitted
and observed intensities, the following relationship holds true:

\begin{equation}
\frac{I_{\omega}}{\omega^{3}n^{2}}=const.
\end{equation}

In vacuum, the frequency experiences Doppler shift $\omega=\gamma\left(\omega^{\prime}+\mathbf{k}^{\prime}\mathbf{v}\right)=\gamma\omega^{\prime}\left(1+\beta\cos\theta^{\prime}\right)$
and the blackbody radiation for a moving observer becomes

\begin{equation}
I_{\omega\prime}^{\prime}=\frac{\omega^{\prime3}}{\omega^{3}}I_{\omega}=\frac{\omega^{\prime3}}{\omega^{3}}I_{bb}=\frac{\hbar}{\pi^{2}c^{2}}\frac{\omega^{\prime3}}{e^{\frac{\hbar\omega\prime}{T^{\prime}}}-1},
\end{equation}

\noindent i.e., for the moving observer the radiation spectrum appears
as blackbody but with new effective directional temperature $T^{\prime}=T/\gamma\left(1+\beta\cos\theta^{\prime}\right)$.

For a moving observer in the plasma universe:

\begin{equation}
I_{\omega\prime}^{\prime}=\frac{\omega^{\prime3}n^{\prime2}}{\omega^{3}n^{2}}I_{\omega}=\frac{\omega^{\prime3}}{\omega^{3}}n^{\prime2}I_{bb}=\frac{\hbar}{\pi^{2}c}\frac{\omega^{\prime3}n^{\prime2}}{e^{\frac{\hbar\omega}{T}}-1}.
\end{equation}

Using the Lorentz transformation we can express $n^{\prime}$ and
$\omega$ in terms of $\omega^{\prime}$ and $\theta^{\prime}$ to
get the expression for $I_{\omega\prime}^{\prime}$ in the moving
frame. We cannot find the analytical expression for a general medium,
but, luckily, for plasma, the dispersion relation is Lorentz invariant:

\begin{equation}
\omega^{\prime2}-k^{\prime2}c^{2}=\omega^{2}-k^{2}c^{2}=\omega_{p}^{2},
\end{equation}

\noindent so that the refractive index for a moving observer has the
same functional dependance as for the stationary observer: $n^{\prime}=ck^{\prime}/\omega^{\prime}=1-\omega_{p}^{2}/\omega^{\prime2}$.
Thus, the intensity measured by the moving observer in plasma is

\begin{equation}
I_{\omega\prime}^{\prime}=\frac{\hbar}{\pi^{2}c}\frac{\omega^{\prime}\left(\omega^{\prime2}-\omega_{p}^{2}\right)}{e^{\frac{\hbar\omega^{\prime}}{T^{\prime}}}-1},
\end{equation}

\noindent i.e., for the moving observer the radiation spectrum appears
as that of equilibrium radiation in plasma but with new frequency-dependent
effective directional temperature $T^{\prime}=T/\gamma\left(1+\beta\sqrt{1-\omega_{p}^{2}/\omega^{\prime2}}\cos\theta^{\prime}\right)$.

In reality most of the plasmas are not in thermal equilibrium. If
we are interested in radiation from the plasma, we should know specific
emission mechanism in plasma and its optical depth. The most prominent
example of the equilibrium radiation in nature is CMB. We are going
to study plasma effects on it in the next section.

\section{Early universe and plasma effects on CMB}

According to accepted cosmology models, the radiation in the early
universe and ionized matter (plasma) were tightly coupled and thermalized
due to Thomson scattering, until, at cosmological redshift $z\approx1100$,
because of the recombination, the radiation became decoupled from
now essentially neutral matter. This radiation from the early universe
is known as CMB.

Experimental data show that CMB spectrum is consistent with Planck's
law to very high accuracy \citep{Fixsen2009}. In principle, however,
since before the recombination the universe was in the plasma state
for which waves with frequencies below the plasma frequency $\omega_{p}$
cannot propagate, the radiation inside it should have been different
from the Planck spectrum. For this reason, in this section we discuss
the influence of plasma effects on CMB and the possibility of the
experimental observation of new effects.

\subsection{Direct detection of plasma spectrum}

A comparison of the equilibrium radiation in plasma given by Eq.~(\ref{eq:E_x})
with experimental data on CMB radiation was investigated in Ref.~\citep{Colafrancesco2015}.
Three methods of the detection of plasma dispersion effects were proposed.
One is the direct observation of the plasma effects, specifically,
the cutoff at plasma frequency, by comparing experimental data on
CMB with the distribution given by Eq.~(\ref{eq:E_x}). The second
is the modification of the Sunyaev-Zel\textquoteright dovich effect
as the plasma modified CMB radiation travels through the electron
gas of galaxy clusters. The third is the modification of the cosmological
21-cm background radiation. The conclusion reached in Ref.~\citep{Colafrancesco2015}
can be summarized as that, even though the currently available experimental
data cannot show any plasma effects, they might be detectable with
the new generation of low-frequency experiments such as SKA-LOW. We
argue that the conclusion reached in Ref.~\citep{Colafrancesco2015}
is too optimistic for two reasons.

First, the values of parameter $a=(3.63,\:6.10,\:7.36)\times10^{-3}$
used in Ref.~\citep{Colafrancesco2015} were estimated through a
numerical fit to COBE-FIRAS data, to some other data in the range
$\sim$$1.3\lyxmathsym{\textendash}50\:\textrm{GHz}$ (see Ref.~\citep{Colafrancesco2015}),
and to three low-frequency data points from Refs.~\citep{Howell1967,Sironi1990,Sironi1991}.
This procedure of obtaining $a$ is likely to significantly overestimate
it, because the value of $a$ is mostly determined by the above mentioned
three low-frequency data points, which, besides having high uncertainty,
can give only the upper limit on $a$. Simply put, in this case the
parameter $a$ is essentially approximated by the lowest experimental
value available, but the absence of low-frequency data should not
determine the plasma frequency, which actually can be much lower than
the lowest experimental value. Indeed, the value of parameter $a$
estimated through the electron density ($n_{e}\approx300\:\textrm{cm}^{-3}$
\citep{Triger2010}) and the CMB temperature ($T/k_{B}\approx3000\:\textrm{K}$)
just before the recombination gives orders of magnitude lower value
$a\approx2\times10^{-9}$. Plasma dispersion brings corrections on
the order of $O\left(a^{2}/x^{2}\right)$ or $O\left(\omega_{p}^{2}/\omega^{2}\right)$,
which is a small number. Indeed, the value of electron density just
before the recombination is $n_{e}\approx300\:\textrm{cm}^{-3}$ \citep{Triger2010},
with corresponding plasma frequency $\omega_{p}\approx10^{6}\:\textrm{\ensuremath{\textrm{s}^{-1}}}$.
The smallest frequency measurable by Planck's spacecraft is $\nu_{min}=30\:\textrm{GHz}$
\citep{Planck2010}, which corresponds to $\omega=2\pi\nu\left(1+z\right)\approx2\times10^{14}\:\textrm{\ensuremath{\textrm{s}^{-1}}}$
at $z\approx1100$, giving $\omega_{p}^{2}/\omega^{2}\approx2\times10^{-17}$.
For the proposed SKA-LOW experiment, the minimum frequency is $\nu_{min}=50\:\textrm{MHz}$
\citep{Dewdney2013}, which corresponds to $\omega=2\pi\nu\left(1+z\right)\approx3\times10^{11}\:\textrm{\ensuremath{\textrm{s}^{-1}}}$
at $z\approx1100$, giving $\omega_{p}^{2}/\omega^{2}\approx8\times10^{-12}$.

Second, even if, before the recombination, the CMB spectrum were given
by Eq.~(\ref{eq:E_x}), it would have been modified significantly
during the recombination. The after-recombination spectrum of CMB
depends on how fast the recombination happened. If it were so slow
that full thermal equilibrium was maintained at every step (this requires
destruction and creation of photons), then, after the recombination,
we would get the radiation spectrum (\ref{eq:E_x}) with the parameter
$a$ equal to zero (for simplicity, we consider complete recombination
into neutral state, in reality the electron density decreases by about
a factor of $10^{3}-10^{4}$ \citep{Sunyaev2009}), which is just
the Planck distribution (maybe with nonzero $\mu$) and no plasma
dispersion effects would be present. If, on the other hand, the recombination
were so sudden that the number of photons is conserved, then the wave
vector (and wavelength) of each radiation mode would remain constant,
while the frequency would change: $k=k_{0}$, $\omega=kc=\sqrt{\omega_{0}^{2}-\omega_{p}^{2}}$.
The number of photons $dn_{\gamma}$ would not change, the total energy
density of photons would decrease (part of the initial energy would
be converted into heat), and the energy spectrum would become

\begin{equation}
u_{\gamma}=\frac{T^{4}}{\pi^{2}\hbar^{3}c^{3}}\int_{0}^{\infty}\frac{x^{3}}{e^{\sqrt{x^{2}+a^{2}}-\mu^{\prime}}-1}dx.\label{eq:spectrum_rec_ad}
\end{equation}

This is consistent with the adiabatic formula from Ref.~\citep{Dodin2010}
that says that the energy density per frequency {[}$du_{\gamma}\left(x\right)/x${]}
remains constant as the density of plasma changes. Adiabatic here
means slow in comparison with period of the wave $d\ln\omega_{p}/dt\ll\omega$
making the number of waves an adiabatic invariant, but not so slow
that the full equilibrium is established. The amount of energy taken
from CMB and converted to heat can be approximated for small values
of $a$ as $\triangle u_{\gamma}/u_{\gamma}=5a^{2}/2\pi^{2}$. Notice,
though, that part of this energy would be radiated back into CMB because
excited recombined atoms would radiate photons as they fall back into
the ground state. The detailed physics of this process is complicated;
the thorough numerical calculations can be found in Ref.~\citep{Chluba2016}.
According to standard cosmology, the active cosmological hydrogen
recombination happened between $z\approx1600$ and $z\approx800$
and the electron density decreased from $1$ to about $10^{-4}-10^{-3}$
\citep{Sunyaev2009}. Thus, the adiabatic scenario should have been
realized.

\begin{figure}
\includegraphics[width=1\columnwidth]{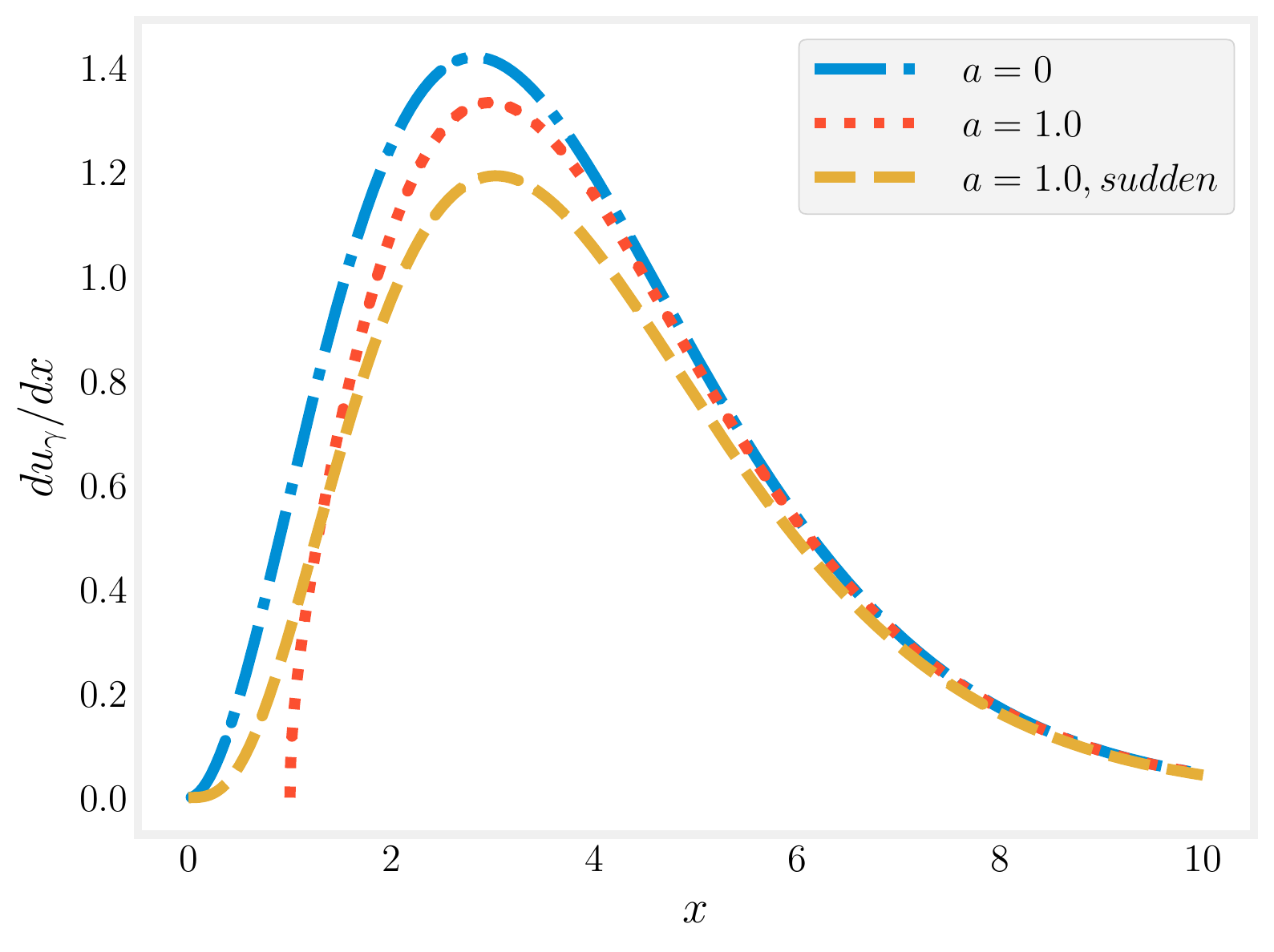} \caption{\label{fig04}Photon spectral energy density in the units of $T^{4}/\pi^{2}\hbar^{3}c^{3}$
versus normalized frequency $x=\hbar\omega/T$ for $a=1$, $\mu=0$
before and after sudden recombination, and for a blackbody of the
same temperature with $a=0,\:\mu=0$. }
\end{figure}

\begin{figure}
\includegraphics[width=1\columnwidth]{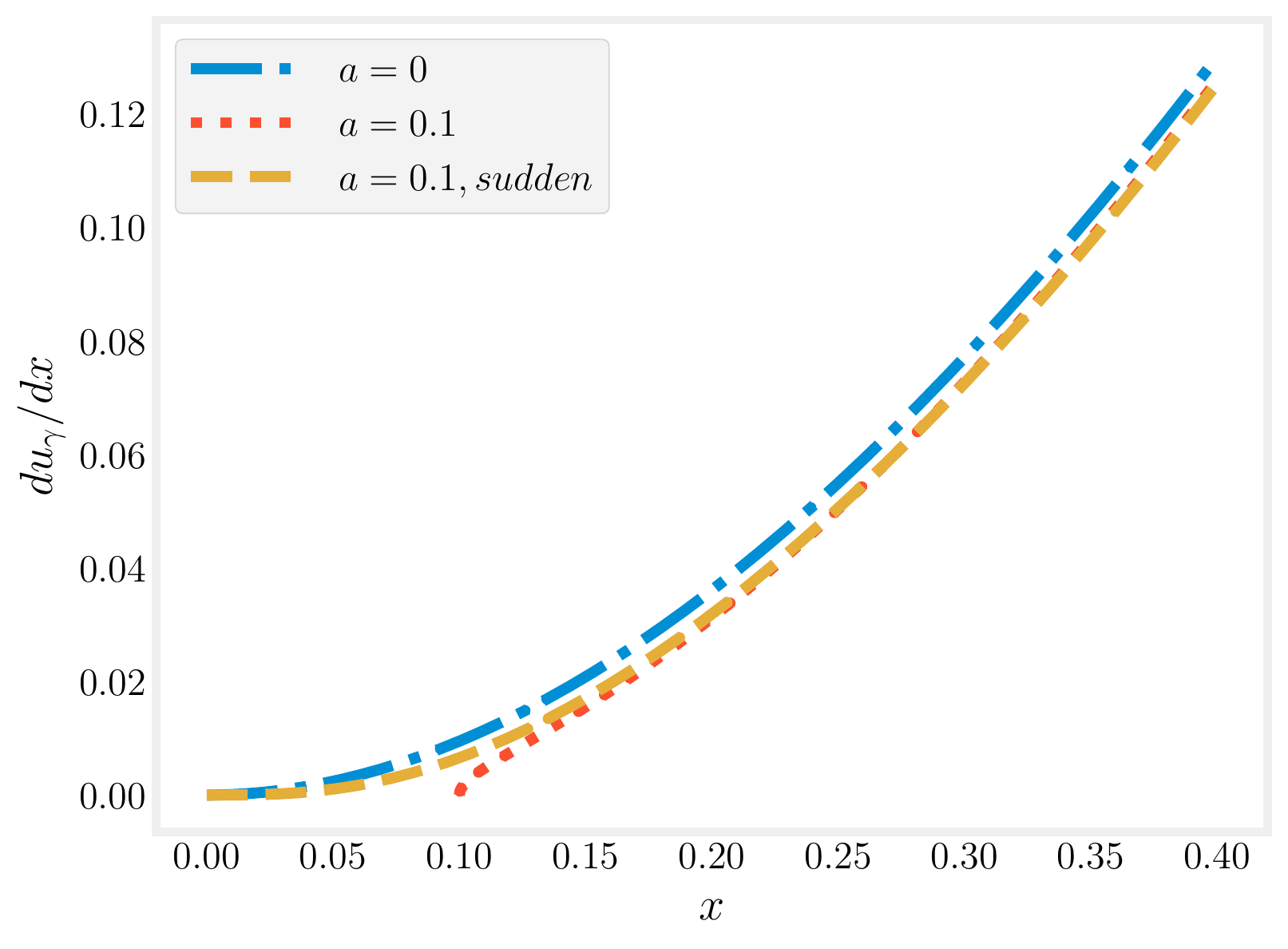} \caption{\label{fig05}Photon spectral energy density in the units of $T^{4}/\pi^{2}\hbar^{3}c^{3}$
versus normalized frequency $x=\hbar\omega/T$ for $a=0.1$, $\mu=0$
before and after sudden recombination, and for a blackbody of the
same temperature with $a=0,\:\mu=0$. }
\end{figure}

Let us compare the spectral energy density given by Eq.~(\ref{eq:spectrum_rec_ad})
with the original spectral energy density in plasma given by Eq.~(\ref{eq:E_x}).
Figure~\ref{fig04} shows the spectral energy density versus the
normalized frequency for $a=1.0$ before and after sudden recombination,
as well as the blackbody spectrum for the same temperature. We see
that the distribution after the sudden recombination resembles the
blackbody spectrum with different temperature. Most importantly, it
does not have cutoff at low frequencies. For small $a$, the difference
between the spectral radiation after the sudden recombination and
that of blackbody is especially hard to notice as demonstrated in
Fig.~\ref{fig05} for $a=0.1$. We remind the reader that the estimate
for parameter $a$ just before the recombination is $a\approx2\times10^{-9}\ll1$.
Since Eq.~(\ref{eq:spectrum_rec_ad}) with $\mu^{\prime}=0$ has
$e^{\sqrt{x^{2}+a^{2}}}-1\approx e^{x+a^{2}/2x}-1$ in the denominator
in contrast to $e^{x}-1$ for the blackbody spectrum, then, for $x\gg a$,
plasma effects can appear as frequency-dependent chemical potential
$\mu_{CMB}=a^{2}/2x$ (see definition of $\mu_{CMB}$ in the next
subsection). However, this chemical potential is extremely small.
Indeed, for $x=2.8$, which corresponds to the maximum of the blackbody
spectrum, $\mu_{CMB}$ has a negligible value of about $7\times10^{-19}$,
for $x_{min}=5\times10^{-1}$ corresponding to the smallest frequency
measured by Planck's spacecraft $\mu_{CMB}\approx4\times10^{-18}$,
and for the SKA-LOW experiment $x_{min}=9\times10^{-4}$ and $\mu_{CMB}\approx2\times10^{-15}$.
For reference, $\mu$-distortion expected from different processes
in $\Lambda$CDM cosmology is about $10^{-9}-10^{-8}$ \citep{Chluba2016b}.
This is itself a very small quantity and has not been experimentally
detected yet, but still, it is about 6 to 10 orders of magnitude higher
than the chemical potential from plasma effect described above.

The recombination was not complete but the electron density decreased
in $10^{4}$ times, which gives parameter $a\approx2\times10^{-11}$
just after the recombination. It is obvious, though, that detection
of the effects related to this, even smaller parameter $a$, is harder.

Thus, the detection of plasma dispersion effects on CMB, such as the
existence of the lower-frequency cutoff, is much harder than suggested
in Ref.~\citep{Colafrancesco2015}. Even if the spectrum of CMB were
indeed given by Eq.~(\ref{eq:E_x}) before the recombination, after
the recombination, the plasma dispersion effects could have been erased
almost completely. 

\subsection{Plasma modification of redshift}

Another apparent plasma effect is change in redshift. The usual cosmological
redshift is described by 
\begin{equation}
\frac{\dot{\omega}}{\omega}=-\frac{\dot{a}_{sc}}{a_{sc}},\label{eq:redshift_st}
\end{equation}
where $a_{sc}\left(t\right)=1/(1+z)$ is the cosmic scale factor and
is not related to parameter $a=\hbar\omega_{p}/T$. In a dispersive
medium Eq.~(\ref{eq:redshift_st}) takes different form \citep{Bicak1975}:

\begin{equation}
\frac{\dot{\omega}}{\omega}=-\left(1+\frac{\omega}{n}\frac{\partial n}{\partial\omega}\right)^{-1}\left(\frac{\dot{a}_{sc}}{a_{sc}}+\frac{\dot{n}}{n}\right),
\end{equation}

\noindent where the time derivative in $\dot{n}$ is not applied to
$\omega$ in $n$. For plasma it gives a frequency-dependent redshift 

\begin{equation}
\frac{\dot{\omega}}{\omega}=-\left(1+\frac{1}{2}\frac{\omega_{p}^{2}}{\omega^{2}}\right)\frac{\dot{a}_{sc}}{a_{sc}},\label{eq:exp_plasma}
\end{equation}

\noindent which is consistent with the expression obtained in Ref.~\citep{Dodin2010b}. 

According to Eq.~(\ref{eq:redshift_st}), frequency scales as $1+z$,
and since temperature also scales as $1+z$, their ratio $x=\hbar\omega/T$
is scale invariant, so that the blackbody radiation spectrum remains
blackbody-shaped as the universe expands. In contrast, according to
Eq.~(\ref{eq:exp_plasma}), the frequency does not scale simply as
$1+z$ anymore, suggesting that in plasma the shape of the spectrum
does not remain the same as the universe expands. Intuitively, it
is because light, which is inherently relativistic, in plasma acquires
some properties of the matter, which is nonrelativistic. For ultrarelativistic
plasma the plasma frequency is $\omega_{p}^{2}=4\pi ne^{2}c^{2}/3T$.
In this case the frequency would grow in the same way as temperature.

It is believed that the universe was in full equilibrium at $z_{0}=2\times10^{6}$,
which defines the blackbody surface \citep{Khatri2012b}. In the absence
of plasma and purely under the influence of cosmological redshift
(no distortions) the blackbody spectrum at $z_{0}$ would be transformed
into another blackbody spectrum at $z=1100$ with the new temperature
$T=T_{0}\left(1+z\right)/\left(1+z_{0}\right)$. Now, with plasma,
the full equilibrium spectrum at $z_{0}$ was given by Eq.~(\ref{eq:E_x})
with zero chemical potential. From Eq.~(\ref{eq:exp_plasma}) we
obtain $\left(\omega^{2}-\omega_{p}^{2}\right)/\left(\omega_{0}^{2}-\omega_{p0}^{2}\right)=\left(1+z\right)^{2}/\left(1+z_{0}\right)^{2}$
\citep{Bicak1975} and, using $\omega_{p}^{2}\propto n_{e}\propto\left(1+z\right)^{3}$,
we can calculate what the full equilibrium spectrum at $z_{0}=2\times10^{6}$
would look like at $z=1100$ under plasma modified cosmological redshift
(ignoring all other processes causing distortion):

\begin{equation}
du_{\gamma}=\frac{T^{4}}{\pi^{2}\hbar^{3}c^{3}}\frac{x\sqrt{x^{2}-a^{2}}\sqrt{x^{2}+\frac{z_{0}-z}{1+z}a^{2}}}{e^{\sqrt{x^{2}+\frac{z_{0}-z}{1+z}a^{2}}}-1}dx.
\end{equation}

We see that while plasma redshift does not change the cutoff frequency
it brings additional corrections on the order of $a^{2}\left(z_{0}-z\right)/\left(1+z\right)\approx a^{2}\left(z_{0}/z\right)\approx4\times10^{-15}$,
i.e., these corrections are determined by parameter $a_{0}=a\left(1+z_{0}\right)^{1/2}/\left(1+z\right)^{1/2}\approx a\left(z_{0}/z\right)^{1/2}$
at $z_{0}=2\times10^{6}$ rather than parameter $a$ at $z=1100$.
In particular it can manifest itself as a frequency-dependent chemical
potential $\mu_{CMB}\approx\left(z_{0}/z\right)a^{2}/2x$. This chemical
potential is about $\left(z_{0}/z\right)\approx10^{3}$ times larger
than the one considered in the previous subsection but it is still
beyond the values that can be experimentally detected. In addition,
we note that for SKA-LOW at its lowest frequency $\nu_{min}=50\:\textrm{MHz}$,
CMB foregrounds are going to further complicate measurements. Thus,
the plasma correction to the redshift is extremely small and can hardly
be detected by any past and near future CMB experiments.

Similar plasma redshift corrections would take place after the recombination
for lower redshifts before (since the recombination is not complete)
and after the reionization, but these corrections are smaller, since
parameter $a$ at $z_{0}=2\times10^{6}$ is higher.

\section{Modification of the Kompaneets equation}

The standard way to describe the thermalization of the radiation and
electrons through Thomson scattering and to quantify distortions of
CMB from blackbody is to use the Kompaneets equation \citep{Kompaneets1957}
($x_{e}=\hbar\omega/T_{e}$):

\begin{equation}
\frac{\partial n_{\gamma}}{\partial t}=\frac{n_{e}\sigma_{T}T_{e}}{m_{e}c}\frac{1}{x_{e}^{2}}\frac{\partial}{\partial x_{e}}\left[x_{e}^{4}\left(\frac{\partial n_{\gamma}}{\partial x_{e}}+n_{\gamma}+n_{\gamma}^{2}\right)\right],\label{eq:Komp}
\end{equation}

\noindent where $\sigma_{T}$ is the Thomson cross section.

The equilibrium solution ($\partial n_{\gamma}/\partial t=0$) of
this equation is given by the Bose-Einstein distribution with, in
general, nonzero chemical potential $\mu$ (we note that it is customary
to use a different sign convention for chemical potential in CMB science
in comparison with the statistical mechanics literature: $\mu_{CMB}=-\mu^{\prime}=-\mu/T$).
The chemical potential appears because Eq.~(\ref{eq:Komp}) accounts
only for scattering and, consequently, conserves the number of photons.
This deviation from the Planck distribution is known as $\mu$-distortion.
If it exists, $\mu$-distortion is very small: according to the COBE-FIRAS
data the constraint on $\mu$ is $\left|\mu\right|/T<9\times10^{-5}$
\citep{Fixsen1996}, while recent Planck data put even stronger constraint:
$\left|\mu\right|/T<6.1\times10^{-6}$ \citep{Khatri2015}. In the
limit of small optical depth for electron Compton scattering, Eq.~(\ref{eq:Komp})
describes another type of distortion called Compton $y$-distortion,
where $dy=n_{e}\sigma_{T}cdt\left(T_{e}-T_{\gamma}\right)/m_{e}c^{2}$
(this $y$ is not related to the normalized wave vector used before).
The constraint on $y$-distortion is also very strong: $\left|y\right|<1.5\times10^{-5}$
according to COBE-FIRAS \citep{Fixsen1996}. The intermediate regime
between the two extremes is called $i$-distortion \citep{Tashiro2014}.

We want the modified version of Eq.~(\ref{eq:Komp}) that includes
the influence of plasma dispersion and has the equilibrium solution
corresponding to the energy density given by Eq.~(\ref{eq:E_x}).
This generalization of Eq.~(\ref{eq:Komp}) to plasma environment
was derived in Ref.~\citep{Mendoca2017}, where the possibility of
Bose-Einstein condensation in scattering dominated plasma was investigated:

\begin{equation}
\frac{\partial n_{\gamma}}{\partial t}=\frac{n_{e}\sigma_{T}T_{e}}{m_{e}c}\frac{1}{y_{e}^{2}}\frac{\partial}{\partial y_{e}}\left[y_{e}^{4}\left(\frac{\sqrt{y_{e}^{2}+a_{e}^{2}}}{y_{e}}\frac{\partial n_{\gamma}}{\partial y_{e}}+n_{\gamma}+n_{\gamma}^{2}\right)\right],\label{eq:Komp_plasma_ye}
\end{equation}

\noindent where $y_{e}=\hbar kc/T_{e}=\sqrt{x_{e}^{2}-a_{e}^{2}}$
and $a_{e}=\hbar\omega_{p}/T_{e}$. 

An interesting feature of Eqs.~(\ref{eq:Komp_plasma_ye}) and~(\ref{eq:Komp})
is that if the temperature of the photons is initially higher than
the electron temperature ($T_{\gamma}>T_{e}$), then the Thomson scattering
leads to the formation of a peak near $k=0$ or $\omega=\omega_{p}$
{[}$\omega=0$ for Eq.~(\ref{eq:Komp}){]}. This pile-up of photons
near low frequencies is reminiscent of Bose-Einstein condensation
\citep{Zeldovich1969,Zeldovich1975,Mendoca2017,Mati2019}. In the
absence of energy injection, the situation with $T_{\gamma}>T_{e}$
is realized for CMB: as the universe expands the temperature of the
radiation scales inversely with the scale factor: $T_{\gamma}\propto\left(1+z\right)$,
while the temperature of the matter scales as $T_{e}\propto\left(1+z\right)^{2}$.
It is then usually argued that the Bose-Einstein condensation does
not actually take place in reality, because at low frequencies absorption
mechanisms, such as Bremsstrahlung and double Compton scattering,
become important. Thus, Thomson scattering redistributes the excess
energy among photons creating large number of low energy photons,
which are then absorbed at low frequency. This leads to a negative
(if $\mu$ is defined in terms of CMB community convention) adiabatic
cooling $\mu$-distortion \citep{Khatri2012,Chluba2016b}, which partly
cancels positive $\mu$-distortion due to energy injection. In plasma,
Bremsstrahlung and double Compton scattering are reduced around plasma
frequency, so some other photon absorption mechanisms should take
place. Since Thomson scattering creates an excess of photons at low
frequencies, the electromagnetic wave energy can significantly exceed
the thermal level, making unique plasma mechanisms of radiation absorption
effective at getting rid of the excess photons at low frequency. For
example, different types of parametric instabilities, such as the
two-plasmon decay, when the electromagnetic wave with frequency around
$2\omega_{p}$ decays into two plasmons. In addition, collisionless
absorption due to Landau damping might be important.

Equation~(\ref{eq:Komp_plasma_ye}) allows one to study the evolution
of the CMB spectrum and its deviation from blackbody radiation taking
into account both Thomson scattering and change in the dispersion
relation of photons in plasma. The resulting distortions, however,
would be significantly different from the ones obtained through Eq.~(\ref{eq:Komp})
only for low frequencies around $\omega\approx\omega_{p}$.

A further step in generalizing Eq.~(\ref{eq:Komp}) can be made by
including the influence of the collective plasma effects on Thomson
scattering. It is known that for wavelengths larger than the Debye
length $\lambda_{D}=v_{th}/\omega_{p}$, where the electron thermal
speed is $v_{th}=T_{e}/m_{e}$, the electromagnetic waves see not
the collection of individual independent electrons but rather correlated
dressed particles, which leads to coherent rather than incoherent
scattering and can reduce the effective scattering cross section \citep{Bingham2003}.
Condition $\lambda=\lambda_{D}$ corresponds to frequency $\omega=\omega_{p}\sqrt{c^{2}/v_{th}^{2}-1}\approx\omega_{p}\left(c/v_{th}\right)$,
which means frequencies $c/v_{th}$ times larger than the ones in
Eq.~(\ref{eq:Komp_plasma_ye}) would be affected by this.

The generalization of Eq.~(\ref{eq:Komp}) that takes into account
the collective effects in Thomson scattering was suggested in Ref.~\citep{Bingham2003}.
According to Ref.~\citep{Bingham2003}, the Kompaneets equation with
the collective effects taken into account (wave time dispersion is
ignored here) can be written as

\begin{equation}
\frac{\partial n_{\gamma}}{\partial y}=\frac{1}{x_{e}^{2}}\frac{\partial}{\partial x_{e}}\left[I_{e}\left(\delta_{e}\right)x_{e}^{4}\left(\frac{\partial n_{\gamma}}{\partial x_{e}}+n_{\gamma}+n_{\gamma}^{2}\right)\right].\label{eq:Komp_collective}
\end{equation}

\noindent Here the collective parameter $\delta_{e}$ is introduced:
\begin{equation}
\delta_{e}=\frac{\omega_{p}^{2}}{\omega^{2}-\omega_{p}^{2}}\frac{c^{2}}{2v_{th}^{2}},
\end{equation}

\noindent and function of the collective parameter $I_{e}\left(\delta_{e}\right)$
is defined through the following integral:

\begin{multline}
I_{e}\left(\delta_{e}\right)=\frac{3}{4}\int_{-1}^{1}d\zeta\left(1+\zeta^{2}\right)\left(1-\zeta\right)^{3}\\
\times\frac{1}{\sqrt{\pi}}\int_{-\infty}^{\infty}\frac{\xi^{2}e^{-\xi^{2}}d\xi}{\left|1-\zeta+\delta_{e}W\left(\xi\right)\right|^{2}},\label{eq:I_e}
\end{multline}

\noindent 
\begin{equation}
W\left(\xi\right)=1-2\xi e^{-\xi^{2}}\int_{0}^{\xi}e^{t^{2}}dt+i\sqrt{\pi}e^{-\xi^{2}}=1+Z\left(\xi\right),
\end{equation}

\noindent where $Z\left(\xi\right)$ is plasma dispersion function
for real arguments.

\begin{figure}
\includegraphics[width=1\columnwidth]{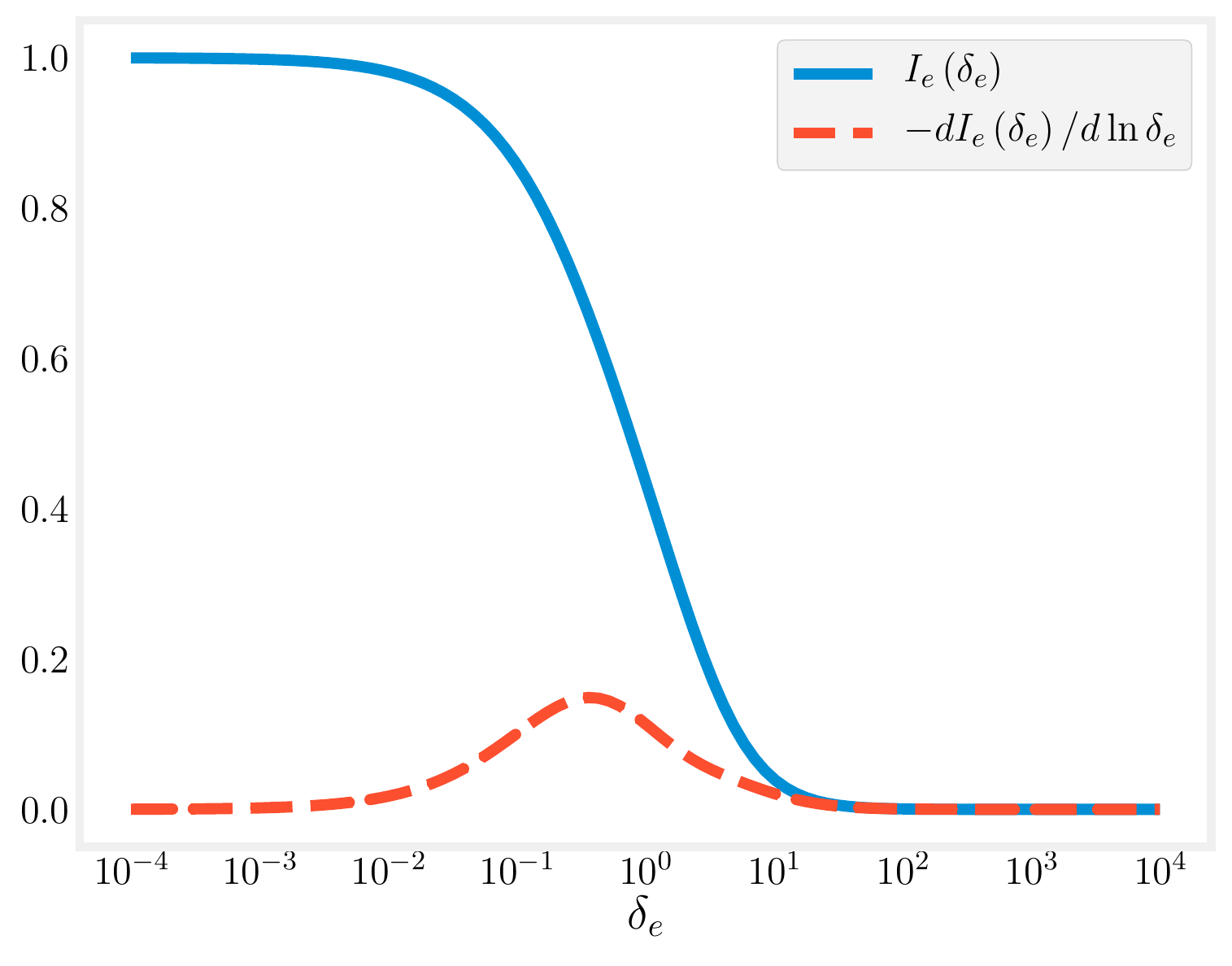} \caption{\label{fig06}Function $I_{e}\left(\delta_{e}\right)$ defined by
Eq.~(\ref{eq:I_e}) and negative of its logarithmic derivative $dI_{e}\left(\delta_{e}\right)/d\ln\delta_{e}$
versus the collective parameter $\delta_{e}$.}
\end{figure}

Figure~\ref{fig06} shows function $I_{e}\left(\delta_{e}\right)$
and negative of its logarithmic derivative $dI_{e}\left(\delta_{e}\right)/d\ln\delta_{e}$
as a function of the collective parameter $\delta_{e}$. We can see
that the function $I_{e}\left(\delta_{e}\right)$ quickly decreases
between $\delta_{e}=10^{-2}$ and $\delta_{e}=10$ dropping from $I_{e}\approx0.98$
at $\delta_{e}=10^{-2}$ to $I_{e}\approx0.04$ at $\delta_{e}=10$.
This means that for $\delta_{e}\gg1$ the collective effects would
reduce the scattering by orders of magnitude. One should keep in mind,
however, that for very large $\delta_{e}$ the ion contribution to
scattering should be taken into account by adding the appropriate
function $I_{i}\left(\delta_{e}\right)$, see Ref.~\citep{Bingham2003}.
Note also that the graph for function $I_{e}\left(\delta_{e}\right)$
and asymptotic formulas in Ref.~\citep{Bingham2003} appear to be
erroneous.

\section{Plasma effects during and after the epoch of reionization}

\subsection{Plasma corrections to $y$-distortion}

After the epoch of recombination, the universe became ionized again
during the epoch of reionization. The epoch of reionization took place
approximately for the redshifts $z\sim6-15$ \citep{Zaroubi2013}.
As CMB light passes through newly ionized intergalactic medium (IGM)
during the epoch of reionization and, later, postreionization, as
it travels through electron plasma of IGM and through plasma of intracluster
medium (ICM) of clusters of galaxies, the CMB spectrum experiences
scattering on hot electrons ($T_{e}\gg T_{\gamma}$), which results
in Compton $y$-distortion. Modern estimates suggest that sky-averaged
nonrelativistic contributions to $y$-distortion from ICM, IGM, and
reionization are given, respectively, by $\left\langle y\right\rangle _{ICM}=1.58\times10^{-6}$,
$\left\langle y\right\rangle _{IGM}=8.9\times10^{-8}$, $\left\langle y\right\rangle _{reion}=9.8\times10^{-8}$
\citep{Hill2015}, making it the largest expected CMB distortion within
the $\Lambda$CDM cosmology \citep{Chluba2016b}. Let us derive and
estimate corrections from plasma dispersion described by Eq.~(\ref{eq:Komp_plasma_ye})
and from the collective effects described by Eq.~(\ref{eq:Komp_collective})
to $y$-type distortion. Note that our analysis below is different
from the one in Ref.~\citep{Colafrancesco2015} where it was assumed
that plasma modified spectrum given by Eq.~(\ref{eq:E_x}) enters
galaxy cluster, and here we assume that blackbody spectrum enters
galaxy cluster and experiences change in dispersion relation or change
in scattering due to the collective effects while inside it, which
leads to the modification of Compton $y$-distortion.

We rewrite Eq.~(\ref{eq:Komp_plasma_ye}) in terms of the normalized
frequency $x=\hbar\omega/T_{\gamma}=\left(T_{e}/T_{\gamma}\right)x_{e}$
and using smallness of $T_{\gamma}/T_{e}\ll1$ and $y\approx n_{e}\sigma_{T}tc\left(T_{e}/m_{e}c^{2}\right)\ll1$
we keep only $\partial n_{\gamma}/\partial x$ term in the brackets:

\begin{multline}
\frac{\partial n_{\gamma}}{\partial y}=\frac{1}{x\sqrt{x^{2}-a^{2}}}\\
\times\frac{\partial}{\partial x}\left\{ \left(x^{2}-a^{2}\right)^{2}\left[\frac{\partial n_{\gamma}}{\partial x}+\frac{T_{\gamma}}{T_{e}}\left(n_{\gamma}+n_{\gamma}^{2}\right)\right]\right\} \\
\approx\frac{1}{x\sqrt{x^{2}-a^{2}}}\frac{\partial}{\partial x}\left[\left(x^{2}-a^{2}\right)^{2}\frac{\partial n_{\gamma}}{\partial x}\right].\label{eq:Komp_plasma_xe}
\end{multline}

Substituting the equilibrium blackbody solution into the right-hand
side, we obtain the estimate for the change in photon distribution:

\begin{equation}
\frac{\Delta n_{\gamma}}{y}=\frac{\sqrt{x^{2}-a^{2}}e^{x}}{\left(e^{x}-1\right)^{2}}\left(\frac{x^{2}-a^{2}}{x}\frac{e^{x}+1}{e^{x}-1}-4\right).\label{eq:deltan_plasma}
\end{equation}

For $a=0$, the above equation gives the usual $y$-type distortion
with the corresponding change in intensity $\Delta I_{\omega}\left(x\right)=x^{3}\Delta n_{\gamma}\left(x\right)I_{0}$,
where $I_{0}=(\hbar/2\pi^{2}c^{2})\left(T^{3}/\hbar^{3}\right)$.
For $a\ll x$, it gives:

\begin{multline}
\frac{\Delta n_{\gamma}}{y}=\frac{xe^{x}}{\left(e^{x}-1\right)^{2}}\left(x\frac{e^{x}+1}{e^{x}-1}-4\right)\\
-\frac{a^{2}}{x^{2}}\frac{xe^{x}}{\left(e^{x}-1\right)^{2}}\left(\frac{3}{2}x\frac{e^{x}+1}{e^{x}-1}-2\right),\label{eq:delta_plasma_small}
\end{multline}

\noindent where the first term is the usual nonrelativistic $y$-distortion
and the second term is $O\left(a^{2}/x^{2}\right)$ order plasma correction
to it. For completeness we note that, in addition, the same order
$O\left(\omega_{p}^{2}/\omega^{2}\right)$ corrections would also
be present due to reflection at the cluster boundary. Equations~(\ref{eq:deltan_plasma})
and~(\ref{eq:delta_plasma_small}) together with $\Delta I_{\omega}\left(x\right)=x^{3}\Delta n_{\gamma}\left(x\right)I_{0}$
give the shape and value of the plasma modified $y$-distortion.

Similarly, from Eq.~(\ref{eq:Komp_collective}) we obtain:

\begin{multline}
\frac{\Delta n_{\gamma}}{y}=\frac{xe^{x}}{\left(e^{x}-1\right)^{2}}\left[I_{e}\left(\delta_{e}\right)\left(x\frac{e^{x}+1}{e^{x}-1}-4\right)\right.\\
\left.+2\frac{x^{2}}{x^{2}-a^{2}}\frac{dI_{e}\left(\delta_{e}\right)}{d\ln\delta_{e}}\right].\label{eq:deltan_collective}
\end{multline}

Based on Fig.~\ref{fig06} we can see that the effect of Eq.~(\ref{eq:deltan_collective})
is the overall reduction of the effective value of parameter $y$
for such $x$ that $\delta_{e}\gg1$. This reduction and the effects
from plasma dispersion are negligible for relevant parameters, however.
Indeed, the ICM of clusters of galaxies have electron plasma with
temperature about $T_{e}\approx10^{7}\:\textrm{K}$ and density $n_{e}\approx10^{-4}-10^{-2}\:\textrm{c\ensuremath{\textrm{m}^{-3}}}$
\citep{Sarazin1992}. We would take $n_{e}\approx10^{-2}\:\textrm{c\ensuremath{\textrm{m}^{-2}}}$
for estimates, which gives $\omega_{p}\approx6\times10^{3}\:\textrm{\ensuremath{\textrm{s}^{-1}}}$.
For Planck's spacecraft $\nu_{min}=30\:\textrm{GHz}$ and $\omega_{p}^{2}/\omega^{2}\approx10^{-15}$,
for SKA-LOW $\nu_{min}=50\:\textrm{MHz}$ and $\omega_{p}^{2}/\omega^{2}\approx4\times10^{-10}$.
As for the collective plasma correction to $y$-distortion, for Planck's
spacecraft we have $\delta_{e}=8\times10^{-16}$ and the correction
estimated as $\left|1-I_{e}\left(\delta_{e}\right)\right|+0.5\left|dI_{e}\left(\delta_{e}\right)/d\ln\delta_{e}\right|$
is approximately $2\times10^{-15}$, for SKA-LOW $\delta_{e}=3\times10^{-10}$
and the correction is about $9\times10^{-10}$. The collective plasma
corrections to $y$-distortion in IGM should be somewhat higher because
of the lower temperature, but lower temperature in turn makes the
total distortion smaller (parameter $y$), making it harder to detect.

Thus, plasma effects would cause extremely small change, at the order
of less than $10^{-10}$, to the already quite small and not yet detected
$y$-distortion ($y\approx10^{-6}$), which makes us conclude that
plasma effects on $y$-type distortion during and after the epoch
of reionization can hardly be detected in the near future. 

A smaller Compton $y$-distortion with $y\approx10^{-10}-10^{-8}$
also took place for redshift between $z=10^{5}$ and $z=10^{3}$ \citep{Khatri2012b}
and the same procedure can be applied to those conditions. It is obvious,
though, that one would get similarly minuscule corrections.

\subsection{Heating and magnetic field generation}

Unlike cosmological recombination, which is mostly a volumetric uniform
process, cosmological reionization is a patchy nonuniform process.
The universe became reionized because of ultraviolet (UV) light coming
from first stars and galaxies (and maybe x-rays coming from quasars)
\citep{McQuinn2016}. This UV light first ionized overdense regions
and then ionization fronts were moving ionizing the rest of IGM \citep{Loeb2011}. 

In Refs.~\citep{Jiang1973,Wilks1988,Qu2018} it was shown that, when
a plane electromagnetic wave of frequency $\omega_{0}$ experiences
sudden ionization, it is transformed into three modes: forward and
backward propagating frequency upshifted ($\omega=\sqrt{\omega_{p}^{2}+\omega_{0}^{2}}$)
waves, and a static magnetic field mode. It is easy to see that, in
case of unpolarized light, there will be no static magnetic field
mode since the magnetic fields generated for each equally possible
configurations will cancel each other, resulting in heat.

Due to Thomson scattering CMB radiation is linearly polarized at the
level of 10\% \citep{Buzzelli2016,Hu1997}. The linear polarization
was first detected by the Degree Angular Scale Interferometer (DASI)
in 2002 \citep{Kovac2002}. The polarization could be formed at the
epoch of recombination \citep{Buzzelli2016,Kaiser1983}, i.e., way
before the epoch of reionization, so that CMB was likely already polarized
just before the ionization. Thus, part of the energy of this linearly
polarized component could have been transformed into the energy of
static magnetic field. Let us estimate the magnitude of this field. 

According to Refs.~\citep{Wilks1988,Qu2018}, the amount of the initial
wave energy converted into the magnetic field due to instantaneous
ionization is given by

\begin{equation}
\frac{1}{2}\frac{\omega_{p}^{2}}{\omega_{p}^{2}+\omega_{0}^{2}}=\frac{1}{2}\frac{a^{2}}{a^{2}+x^{2}}\approx\frac{1}{2}\frac{a^{2}}{x^{2}},
\end{equation}

\noindent so that the amount of CMB energy density converted into
the magnetic field energy density can be estimated as

\begin{equation}
\frac{B^{2}}{8\pi}\approx0.1\frac{T^{4}}{\pi^{2}\hbar^{3}c^{3}}\int_{0}^{\infty}\frac{a^{2}}{2x^{2}}\frac{x^{3}}{e^{x}-1}dx=\frac{a^{2}}{8\pi^{2}}\frac{4}{c}\sigma_{SB}\left(\frac{T}{k_{B}}\right)^{4}.\label{eq:magnetic_energy}
\end{equation}

The CMB temperature for $z\sim6-15$ was approximately $T/k_{B}\sim19-44\:{\textstyle \textrm{K}}$.
Taking $T/k_{B}\approx20\:\textrm{K}$ and the parameter $a\approx2\times10^{-10}$
for this temperature and $n_{e}\approx10^{-4}\:\textrm{c\ensuremath{\textrm{m}^{-3}}}$
\citep{Inoue2004}, we get that, even though a very small fraction
$a^{2}/8\pi^{2}\approx5\times10^{-22}$ of the original CMB energy
density goes into the magnetic field energy, we still get cosmologically
very large estimates: the magnetic field energy density is at the
order of $10^{-30}\:\textrm{erg\ensuremath{\cdot}c\ensuremath{\textrm{m}^{-3}}}$
and the corresponding magnetic field is about $B\approx10^{-15}\:\textrm{G}$.
For comparison, in Ref.~\citep{Gnedin2000} the magnetic field produced
during reionization by the Biermann battery effect was estimated to
be $B\approx10^{-19}\:\textrm{G}$. Thus, it seems that the presented
mechanism could have been a reason for the origin of the still unexplained
initial magnetic field seed  \citep{Widrow2002,Munirov2017a}. However,
when we take into account the finite time of ionization $\tau_{ion}$,
we find that the ionization happens in the adiabatic regime ($\omega\gg\tau_{ion}^{-1}$),
so most of this energy is converted into heat instead of magnetic
field. The physical picture is that, in case of sudden ionization,
electrons start oscillating approximately at the same phase resulting
in the directed motion, electric current, and magnetic field, while
for adiabatic ionization electrons have random phases resulting in
random motion and heat. The decrease of the magnetic energy with the
growth of the ionization time and its conversion into the heat is
confirmed by numerical simulations \citep{Wilks1988}. 

The ionization time depends on the thickness of the ionization front
and can be estimated as $\tau_{ion}=d/v_{front}$, where $d$ is the
ionization front thickness and $v_{front}$ is the ionization front
speed. According to Ref.~\citep{Deparis2019}, the front speed is
about $v_{front}\sim\left(0.05-0.1\right)c$. The width of the ionization
front can be estimated to be several mean free paths of UV photons
\citep{Cantalupo2011,Lee2016}. The mean free path is around $\lambda_{mfp}\sim1$
physical kpc. The amount of energy going into the magnetic field exponentially
decreases with the growth of the ionization time $\tau_{ion}$ as
$\propto e^{-\tau_{ion}\omega}.$ Then, even for low frequencies around
plasma frequency $\omega_{p}\approx5\times10^{2}\:\textrm{\ensuremath{\textrm{s}^{-1}}}$,
$\tau_{ion}\omega_{p}\gg1$ and the attenuation factor is practically
zero $e^{-\tau_{ion}\omega_{p}}\approx0$, the magnetic field is practically
zero, too, with all the energy going into heat. The total amount going
into heat is higher by about one order of magnitude if we account
for the unpolarized part of CMB light and is roughly $a^{2}/\pi^{2}\approx10^{-21}$,
which gives a tiny temperature increase of about $\triangle T/k_{B}\approx10^{-20}\:{\textstyle \textrm{K}}$.

The possibility of the generation of seed magnetic field should not
be completely ruled out, however. For example, regions with higher
than average density and, consequently, higher parameter $a$ can
be present or sharper ionization fronts can be potentially produced.

\section{Summary and conclusions}

We derived, using three different approaches, the equilibrium radiation
spectrum inside plasma. We demonstrated that, because of dispersion,
it is different from the blackbody Planck spectrum. We considered
what stationary and moving observers sent into the plasma-filled universe
would measure and how these results should be interpreted. 

We then considered how plasma can affect the spectrum of CMB radiation.
Namely, we pointed out the change in the cosmological redshift that
can appear as an effective frequency-dependent chemical potential;
we discussed the possibility of the direct detection of plasma equilibrium
spectrum from CMB data, emphasized its sensitivity to how fast the
cosmological recombination happened and reached more pessimistic conclusion
about its experimental detection than before \citep{Colafrancesco2015};
we gave expressions for the modified Kompaneets equation due to plasma
dispersion and collective effects; we calculated how Compton $y$-distortion
during and after the epoch of reionization is changed because of plasma;
and we proposed, estimated, and deemed likely unrealistic a novel
mechanism of magnetic field generation during the epoch of reionization
due to conversion of some of the energy of CMB into magnetic field
energy. 

We concluded that plasma effects are extremely small, on the order
of $O\left(\omega_{p}^{2}/\omega^{2}\right)$ in most cases and hence
cannot be realistically detected in the near future. However, if some
of the assumptions employed here were violated, then there would be
possibilities for larger effects. Thus, it is important to know how
plasma affects the CMB spectrum in order to fully understand the cosmological
evolution of our universe; for example, restrictions imposed by plasma
effects might, in principle, be used to test alternative models of
cosmology as they put boundaries on the baryon density at different
epochs. Moreover, the analysis conducted in the paper could be relevant
to other astrophysical and laboratory situations where radiation interacts
with plasma.

\section*{Acknowledgements}

This work is supported by research Grant No. DOE NNSA DE-NA0003871.

\end{document}